\documentclass[prl, reprint,nolongbibliography,aps,showpacs,superscriptaddress,amsmath,amssymb,floatfix,showkeys
]{revtex4-2}
\RequirePackage{amssymb,latexsym}
\usepackage{graphicx}
\usepackage{mathrsfs}
\usepackage{bm}
\usepackage{color}
\usepackage{orcidlink}
\usepackage{subfigure}
\usepackage{comment}
\usepackage{hyperref}

\newcommand{\cnote}[2]{{\color{#1} #2}}

\newcommand{\del}{\partial}
\newcommand{\eps}{\epsilon}

\begin{document}

\title{Non-linear stability of the matter dominated universe}

\author{David Fajman \orcidlink{0000-0003-3034-6232}}
\affiliation{Faculty of Physics, University of Vienna,
Austria.}
\email{david.fajman@univie.ac.at}
\author{Elliot Marshall \orcidlink{0000-0002-3693-7554}}
\affiliation{Institute of Applied and Computational Mathematics, FORTH, Heraklion, Greece}


\begin{abstract}
We numerically study non-linear perturbations of the Einstein-de Sitter spacetime as a solution to the Gowdy-symmetric Einstein-Euler system for a polytropic equation of state. The results suggest that the Einstein-de Sitter spacetime is stable for sufficiently small but otherwise generic perturbations. This is in stark contrast to the well known instability of this spacetime when the matter model is dust. Moreover, this indicates a previously unknown stable regime of the Einstein-Euler equations with direct implications for cosmology.  
\end{abstract}


\maketitle

\section{Introduction}
The matter-dominated epoch is the longest period in the evolution of the universe, lasting from redshift $z\approx 3500$ until $z\approx 0.4$ \cite{Ryden_2016}. In relativistic cosmology, this regime is typically modelled by the  Einstein-de Sitter spacetime \cite{EinsteindeSitter1932}, also known as the \textit{matter dominated universe}, which is the Friedman-Lemaître-Robertson-Walker (FLRW) solution with vanishing spatial curvature and dust as a matter source. In cosmic time it takes the form
\begin{equation}
\label{eqn:EdS}
g_{\mathrm{EdS}}=-dt^2+t^{\frac{4}{3}}(dx^2+dy^2+dz^2)
\end{equation}
with spatial topology $\mathbb T^3$ or $\mathbb R^3$. It is well known that \eqref{eqn:EdS} is dynamically unstable as a solution to the Einstein-dust system, in the sense that the future asymptotic behaviour of solutions arising from arbitrarily small perturbations of \eqref{eqn:EdS} is determined by growing density perturbations and hence deviates from \eqref{eqn:EdS} \cite{Ryden_2016,Marshall:2025}. Indeed, this instability is often invoked as an explanation for the formation of large-scale structures during the matter-dominated epoch \cite{Ryden_2016}. On the other hand, any model which is used to describe a significant portion of the cosmological evolution should, in principle, be dynamically \textit{stable} to allow for small inhomogeneities without strongly deviating from the homogeneous geometry over time. From this perspective, it is natural to look for cosmological models which display homogenizing behaviour for small perturbations as is well established for a positive cosmological constant \cite{Friedrich:1986,ringstrom2013,RodnianskiSpeck:2013,Speck:2012,Oliynyk:CMP_2016,LubbeKroon:2013} or curvature domination \cite{Andersson_Moncrief_2011,Fajman_et_al:2021b,FOOW:2023}.

Unlike spacetimes which are curvature or $\Lambda$-dominated, the expansion of matter-dominated  cosmological models is caused by the energy density and pressure of the matter. That is to say, the dynamical properties of the matter model greatly affect the evolution of curvature inhomogeneities in this class of solutions \cite{Taylor:2024,Bernhardt:2025,Marshall:2025}. Thus, in order for such a model to be an attractor, one requires a stabilizing mechanism intrinsic to the matter dynamics, which homogenizes the matter fields and allows for the coupled dynamics to stabilize. Importantly, dust does \textit{not} have such a mechanism due to the decoupling of energy density and velocity.\\
\indent In this paper, we demonstrate that adding an arbitrarily small but non-vanishing pressure to the fluid, by considering a polytropic equation of state with minimal polytropic index $n>3$, stabilizes the matter dominated universe. In particular, this explains the homogenization of spacetime on large scales as a direct consequence of the non-linear interaction of the Einstein-fluid dynamics.\\ 
\indent The polytropic fluid model is characterized by the equation of state $p=K\rho^{1+\frac{1}{n}}$, where $K>0$ with the polytropic index $n>0$. For these fluids, the speed of sound, $\frac{\partial p}{\partial \rho}=(1+\frac1n)K\rho^{1/n}$, decays as the universe expands. This means polytropic fluids have a dust-like pressure for late-times. Moreover, the Einstein-Euler system for a polytropic fluid admits spatially flat FLRW solutions \eqref{eqn:FLRW_Solution_Asymptotic_Gowdy} which asymptote towards $g_{EdS}$ for large times.\\ 
\indent However, in contrast to dust, the pressure of a polytropic fluid is non-vanishing for a positive energy density. Our results suggest that the coupling of energy density and fluid velocity via the pressure term introduces a \textit{balancing mechanism} which dominates the fluid dynamics, particularly in the small data regime (cf.\eqref{eq:energy}). The late-time asymptotics of the energy density, which assure the matter dominated expansion rate, in combination with the homogenization due to the non-trivial pressure, makes the asymptotic behaviour of the fluid compatible with the Einstein-de Sitter asymptotics. This appears to have significant consequences for the dynamics when coupled with the Einstein equations as we demonstrate numerically below.\\ 
\indent
Specifically, we find convincing evidence that for sufficiently small, but otherwise generic, initial inhomogeneities of the metric and fluid variables, the future asymptotic behaviour of the corresponding solutions to the Einstein-Euler system all converge to an Einstein-de Sitter Universe. This indicates the orbital stability of the EdS spacetime \eqref{eqn:EdS} \footnote{The expansion-normalized spatial metric converges to a flat torus near the standard one, which reflects the non-trivial moduli space of flat spatial metrics compatible with \eqref{eqn:EdS}}. In particular, this is the first evidence for the existence of a stable, fluid-filled, cosmological model for the matter dominated epoch and the first stable solution to the Einstein-Euler equations in the decelerated regime. Our results also reveal a mechanism by which cosmological evolution drives the spatial geometry toward a flat metric, offering a mathematical explanation for the emergence of large-scale spatial flatness.
\section{Equations and Numerical Setup}
We numerically study non-linear perturbations of the FLRW solution to the Einstein-Euler equations for a polytropic fluid in the expanding direction of spacetime. In particular, we restrict our attention to $\mathbb{T}^{3}$ Gowdy symmetric spacetimes (cf. \cite{gowdy1974, LeFlochRendall:2011}) of the form
\begin{align*}
g= e^{2(\nu-U)}(-dt^{2} + e^{-2\alpha}dx^{2}) &+ e^{2U}(dy+Adz)^{2} \nonumber \\
&+ e^{-2U+2t}dz^{2}
\end{align*}
where $t$ is an areal time coordinate, see \cite{Chrusciel:1990} for details. {In this case, the metric coefficient functions $\nu$, $U$, $\alpha$ and $A$ are functions of $t\in [t_0,\infty)$ and $x\in[0,2\pi)$ {alone, and the Einstein-Euler equations reduce to a $1+1$ dimensional system.} In Gowdy symmetry, the normalised fluid four-velocity has only two non-zero components, $u=u_0dt+u_1dx$, where $u_1$ depends on $(t,x)\in [t_0,\infty)\times [0,2\pi)$ and $(u^{0})^{2} - e^{-2\alpha}(u^{1})^{2} = e^{-2(\nu-U)}$.
 
We introduce three effective variables for the matter, the scalar velocity
$v = \frac{u^{1}}{e^{\alpha}u^{0}}$, $|v| \in [0,1)$, $\Gamma = \frac{1}{\sqrt{1-v^{2}}}$ and the modified density $\mu = e^{2(\nu-U)}\rho$. The Gowdy symmetric Einstein-Euler equations can be expressed as a system of balance laws, using $\partial_tA=A_0$, $\partial_t U=U_0$,
\begin{align}
\label{eqn:A0_FC_evo}
    &\partial_{t}(e^{-\alpha}A_{0}) - \partial_{x}(A_{1}) = e^{-\alpha}\Big( 4(A_{1}U_{1} - A_{0}U_{0}) + A_{0}\Big), \\
\label{eqn:A1_FC_evo}
    &\partial_{t}(e^{-\alpha}A_{1}) - \partial_{x}(A_{0}) = 0 , \\
\label{eqn:U0_FC_evo}
    &\partial_{t}(e^{-\alpha}(U_{0}-\frac{1}{2})) - \partial_{x}(U_{1}) = e^{-\alpha}\Big(\frac{1}{2}e^{4U-2t}(A_{0}^{2} \nonumber \\
    &- A_{1}^{2}) + \frac{1}{2} - U_{0}\Big), \\
\label{eqn:U1_FC_evo}
    &\partial_{t}(e^{-\alpha}U_{1})  - \partial_{x}(U_{0}) = 0 , \\
\label{eqn:Hamiltonian_Constraint}
&\partial_{t}\nu = 1 + (\Gamma^{2}-1)\mu + e^{\frac{-2}{n}(\nu-U)}K\Gamma^{2}\mu^{\frac{n+1}{n}} \nonumber \\
&+ e^{2\alpha}U_{x}^{2} + U_{t}^{2} + \frac{1}{4}e^{4U-2t}(e^{2\alpha}A_{x}^{2} + A_{t}^{2}), \\
\label{eqn:alpha_evo}
&\partial_{t}\alpha = 1 - \big(\mu - e^{\frac{-2}{n}(\nu-U)}K\mu^{\frac{n+1}{n}}\big),  \\
\label{eqn:Euler_FC_1}
    &\partial_{t}(e^{-\alpha}\tau ) + \partial_{x}(S)  = \frac{1}{4}e^{-\alpha -2t}(\mu - K e^{\frac{-2}{n}(\nu-U)}\mu^{\frac{n+1}{n}})\nonumber \\
    &\Big[e^{4U}(A_{1}^{2} + A_{0}^{2}) + 4e^{2t}\big(U_{1}^{2} + U_{0}(U_{0}-1)\big)\Big]\nonumber \\
    &- Ke^{\frac{-2}{n}(\nu-U)-\alpha}\mu^{\frac{n+1}{n}}, \\
\label{eqn:Euler_FC_2}
    &\partial_{t}(e^{-\alpha}S) + \partial_{x}(H ) = -\frac{1}{2}e^{-\alpha}\big(\mu - Ke^{\frac{-2}{n}(\nu-U)}\mu^{\frac{n+1}{n}}\big) \nonumber \\
    &\Big(e^{4U-2t}A_{1}A_{0} + 2U_{1}(2U_{0}-1)\Big),
\end{align} 
and constraints
\begin{align}
    \label{eqn:A1_constraint}
    0 = C_{A_{1}} &:= A_{1} - e^{\alpha}\partial_{x}A, \\
\label{eqn:U1_constraint}
    0 = C_{U_{1}} &:= U_{1} - e^{\alpha}\partial_{x}U, \\
\label{eqn:Momentum_Constraint}
\partial_{x}\nu &= -e^{-\alpha}v\Gamma^{2}(\mu + e^{\frac{-2}{n}(\nu-U)}K\mu^{\frac{n+1}{n}}) \nonumber \\
&+ \frac{1}{2}e^{4U-2t}A_{x}A_{t} + 2U_{x}U_{t},
\end{align}
where
\begin{align*}
    A_{1} &:= e^{\alpha}A_{x}, \quad U_{1} := e^{\alpha}U_{x}, \\
    \tau &:= (\mu + Ke^{\frac{-2}{n}(\nu-U)}\mu^{\frac{n+1}{n}})\Gamma^{2} \nonumber \\
    &- Ke^{\frac{-2}{n}(\nu-U)}\mu^{\frac{n+1}{n}}, \\
    S &:= (\mu + Ke^{\frac{-2}{n}(\nu-U)}\mu^{\frac{n+1}{n}})\Gamma^{2}v, \\ 
    H &:= (\mu + Ke^{\frac{-2}{n}(\nu-U)}\mu^{\frac{n+1}{n}})\Gamma^{2}v^{2} \nonumber \\
    &+ Ke^{\frac{-2}{n}(\nu-U)}\mu^{\frac{n+1}{n}}.   
\end{align*}

The spatial computational domain is taken to be $[0,2\pi)$ with periodic boundary conditions and $N$ cells. We numerically solve the system \eqref{eqn:A0_FC_evo}-\eqref{eqn:Euler_FC_2} using a local Lax-Friedrichs finite volume discretisation with fifth-order WENO reconstruction and fourth-order Runge-Kutta timestepping. For computational simplicity, we have treated point values and cell averages as equivalent in our code, which is a second-order accurate approximation. Thus, our numerical scheme is formally second-order accurate for smooth solutions. Further details on our code, including recovery of the primitive variables and convergence tests, are given in the supplemental material. \\
\indent The FLRW solution to \eqref{eqn:A0_FC_evo}-\eqref{eqn:Momentum_Constraint} is given by 
\begin{equation}
\begin{aligned}
\label{eqn:FLRW_Solution_Areal}
    g &= \frac{-3}{4}e^{\frac{3t}{2}}(1-Ke^{\frac{-3t}{2n}})^{n}dt^{2} + e^{t}(dx^{2} + dy^{2} + dz^{2}) , \\
    \rho &= e^{\frac{-3t}{2}}(1-Ke^{\frac{-3t}{2n}})^{-n}, u = -\sqrt{\frac{3}{4}}e^{\frac{3}{4}t}(1-Ke^{\frac{-3t}{2n}})^{\frac{n}{2}} dt
\end{aligned}
\end{equation} 
As discussed above, for large $t$ this asymptotes to the standard Einstein-de Sitter metric \eqref{eqn:EdS} in areal coordinates,
\begin{equation}
\begin{aligned}
\label{eqn:FLRW_Solution_Asymptotic_Gowdy}
    g_{\mathrm{EdS}} &= -\frac{3}{4}e^{\frac{3t}{2}}dt^{2} + e^{t}(dx^{2} + dy^{2} + dz^{2}) , \\
    \rho_{\mathrm{EdS}} &= e^{\frac{-3t}{2}},  \,  u_{\mathrm{EdS}} = -\left(\frac{3}{4}\right)^{\frac{1}{2}}e^{\frac{3}{4}t} dt.
\end{aligned}
\end{equation}

\subsubsection{Stable Polytropic Solutions}
Ultimately, we are interested in the asymptotic behaviour of small non-linear perturbations of \eqref{eqn:FLRW_Solution_Areal}. To this end, we have studied classes of initial data at $t=0$ characterised by a single parameter $\eps > 0$ such that $\eps=0$ corresponds to \textit{exact} FLRW initial data. That is to say, $\eps$ parameterises the size, with respect to appropriate Sobolev norms, of our non-linear perturbations. Independent of the specific class of initial data we observe the same universal dynamical behaviour of perturbations of metric and fluid variables away from the background. In this paper, we will focus on the following family of initial data at $t=0$, 
\begin{equation}
\begin{gathered}
\label{eqn:ID_Family_1}
    \alpha = \log\Big(\frac{\sqrt{3}}{2}(1-K)^{\frac{n}{2}}\Big) + \eps, \\
    \nu = \log\Big(\frac{\sqrt{3}}{2}(1-K)^{\frac{n}{2}}\Big) + \eps\cos(x), \\
    U = \eps\cos(x), \quad
    U_{0} = \frac{1}{2}, \quad 
    U_{1} = e^{\alpha}\del_{x}U, \\
    A = \eps\sin(x), \quad
    A_{0} = 0 \quad 
    A_{1} = e^{\alpha}\del_{x}A, \\ 
    v = 0, \quad
    \mu = \frac{3}{4} + \eps\sin(x),       
\end{gathered}
\end{equation}
however we emphasise that the behaviour is qualitatively the same for all initial data sets we have studied.\\ 
\indent
For given fixed $\eps$ and $K$, we find there are values of $n$ such that the fluid variables rapidly homogenize and decay to the background FLRW solution. An example of this behaviour is shown in Figure \ref{fig:Fluid_H1_norms}, where the $\dot H^1$-norm is defined by $\|f\|_{\dot{H}^{1}} := \|\del_{x}f\|_{L^{2}}$. Similarly, the metric functions $A$, $U$, and $\nu$ all homogenize while the function $\alpha$ approaches a fixed spatial profile at late times, see Figure \ref{fig:Metric_H1_norms}. Although the $H^{1}$ norm of $\alpha$ does not decay, this is merely a coordinate effect as we discuss below. To further illustrate that the spacetime does in fact homogenize towards the future, we compute the spatial Ricci and Kretschmann scalars, displayed in Figure \ref{fig:SpatialCurvature_Invariant_Decay}. In particular, the decay of the Kretschmann scalar demonstrates the solution evolves towards a flat spatial geometry. In fact, an expansion-normalized spatial limit metric of the form $e^{2(\nu_\infty-U_\infty)}( e^{-2\alpha_{\infty}(x)}dx^{2})+ e^{2U_\infty}(dy+A_\infty dz)^{2}+ e^{-2U_\infty}dz^{2}$, where $U_\infty,\nu_\infty,A_\infty\in\mathbb R$ and $\alpha_\infty(x)$ is a function, is a flat torus and hence homogeneous.

\begin{figure}[htbp!]
	\includegraphics[width=0.8\columnwidth]{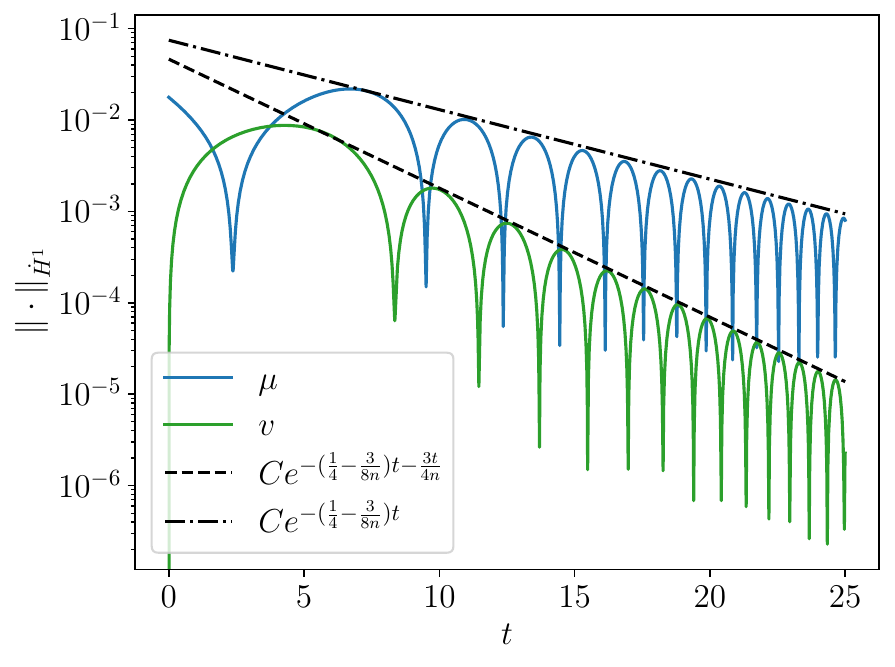}
	\caption{Asymptotic behaviour of the $\dot{H}^{1}$ norms of $\mu$ and $v$. $n=5$, $K=0.2$. The black lines show the theoretical decay rates predicted by the decay estimate \eqref{eqn:decay_estimate_fluid_norms}.}
\label{fig:Fluid_H1_norms}
\end{figure}

\begin{figure}[htbp!]
	\includegraphics[width=0.8\columnwidth]{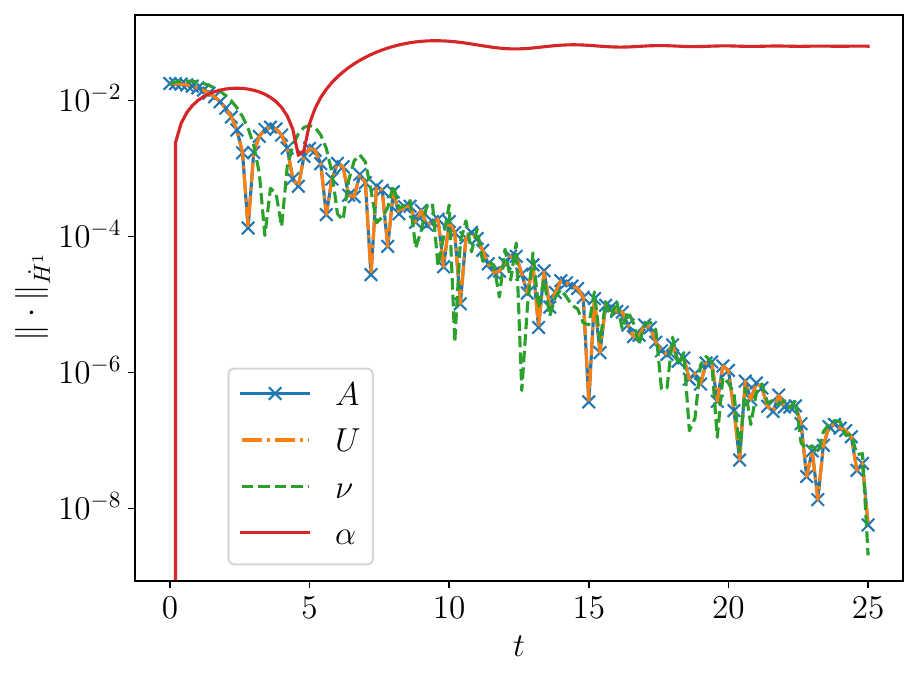}
	\caption{Asymptotic behaviour of the $\dot{H}^{1}$ norms of the metric variables. $n=5$, $K=0.2$, $\eps = 0.01$. We have only displayed a subset of timesteps to improve the clarity of the plot.}
\label{fig:Metric_H1_norms}
\end{figure}

\begin{figure}[htbp!]
	\includegraphics[width=0.8\columnwidth]{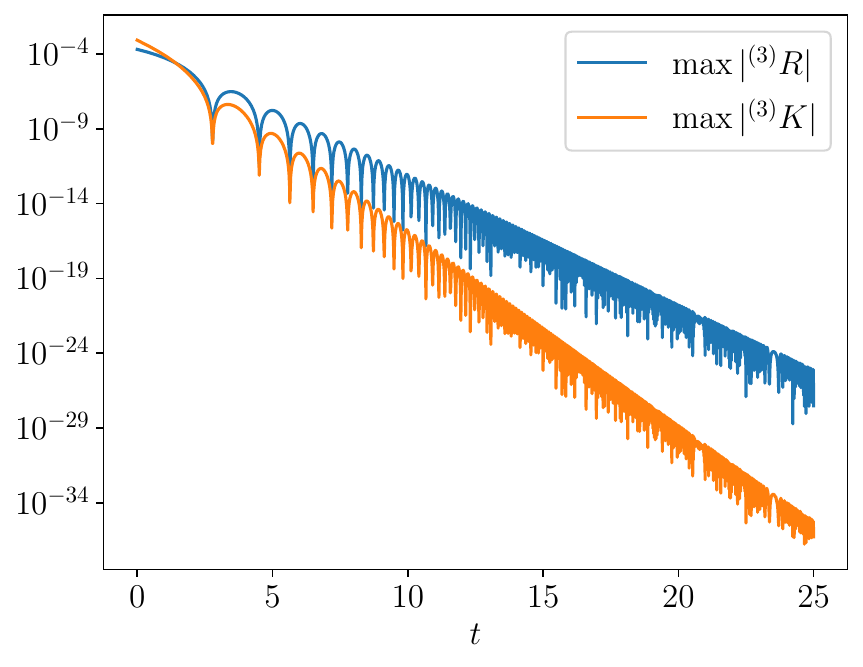}
	\caption{Maximum of the absolute value of the spatial Ricci and Kretschmann scalars. $n = 5$, $K=0.2$, $\eps = 0.01$.
	}
\label{fig:SpatialCurvature_Invariant_Decay}
\end{figure}

The above numerical results suggest that the fluid pressure has a stabilizing effect on the spacetime. Indeed, for all $K>0$ and suitably small $\eps$, we observe stable asymptotics for the polytropic fluid while dust develops shocks for arbitrarily small initial data, see Figure \ref{fig:Dust_Comparison}. This is particularly remarkable given the fact that a polytropic fluid pressure asymptotically approaches dust towards the future as $\rho \searrow 0$. In particular, this implies that a polytropic fluid with arbitrarily small pressure coupled to Einstein's equations is sufficient to stabilize the matter dominated universe.
\begin{figure}[htbp!]
	\includegraphics[width=0.8\columnwidth]{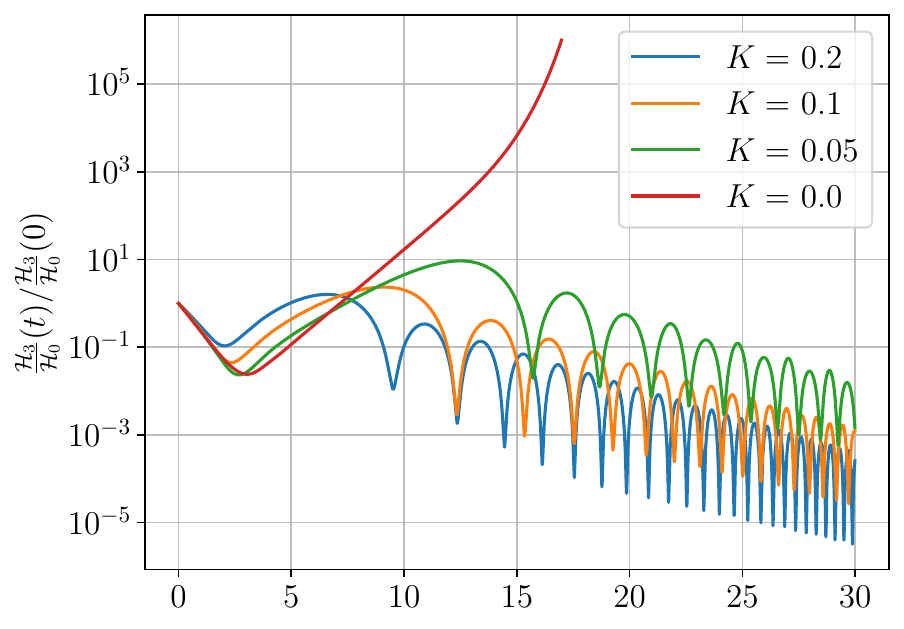}
	\caption{The behaviour of the norm ratio \eqref{eqn:shock_threshold} for different values of $K$. In this example, the norm ratio is computed using $\mu$ instead of $w$ since the latter is undefined when $K=0$. All simulations used $\eps = 0.01, n=5$.}
\label{fig:Dust_Comparison}
\end{figure}
\subsubsection{Transition Between Shock Formation and Stability}
The results of the previous section demonstrate that the FLRW solution is stable for certain fixed values of $\eps$ and $n$. In this section, we analyse the transition from stable to unstable behaviour for varying $\eps$ and $n$.  In the vein of our previous work on isothermal fluids \cite{Fajman_et_al:2024,Marshall:2025}, we monitor the energy functional
\begin{align}\label{H_k-norm}
    \mathcal{H}_{k}(t) := \|\del^{k}_{x}w(t)\|^{2}_{L^{2}} + \|\del^{k}_{x}v(t)\|^{2}_{L^{2}}
\end{align}
where $w := \sqrt{4n(n+1)K}\rho^{\frac{1}{2n}}$ is the so-called Makino density variable. We classify a solution as \textit{stable} if
\begin{align}
\label{eqn:stable_threshold}
    \mathcal{H}_{3}(t) \leq 10^{-6}\mathcal{H}_{3}(0)
\end{align}
and \textit{unstable} if 
\begin{align}
\label{eqn:shock_threshold}
    \left(\frac{\mathcal{H}_{3}}{\mathcal{H}_{0}}\right)(t) \geq 10^{6}\left(\frac{\mathcal{H}_{3}}{\mathcal{H}_{0}}\right)(0)
\end{align}
 Our numerical results indicate that for suitably small, fixed values of $\eps$ and $K$, there exists a critical value $n^{*}$ of the polytropic index such that solutions with $n > n^{*}$ display stable behaviour and solutions with $n < n^{*}$ develop shocks. An example of this transition for $\eps = 0.01$ is shown in Figure \ref{fig:Shock_Stable_Tranisition}. In order to determine the value of $n^{*}$, we perform a scaling analysis by varying the polytropic index and measuring the time of reaching the stabilization or shock thresholds \eqref{eqn:stable_threshold}-\eqref{eqn:shock_threshold}. For $n$ suitably close to $n^{*}$, we find the shock and stabilization times scale like
\begin{equation}
\begin{aligned}
\label{eqn:Shock_Stable_Asymptotic_Scaling}
t_{\text{Shock}} &\approx -a_{1}\log(n^{*}-n) + b_{1}, \\ 
t_{\text{Stable}} &\approx -a_{2}\log(n-n^{*}) + b_{2}
\end{aligned}
\end{equation}
for positive real constants $a_{i}$ and $b_{i}$. For a given fixed $\eps$, the corresponding value of $n^{*}$ is estimated by fitting the threshold times to \eqref{eqn:Shock_Stable_Asymptotic_Scaling}. By repeating this analysis for varying $\eps$, we observe that $n^{*}$ decreases monotonically as $\eps \searrow 0$. This is consistent with the rigorous analysis in the following section; as the size of the initial perturbation is decreased the estimated value of $n^{*}$ approaches approximately $3.10$, while the theoretical lower bound for stable solutions is $n>3$.   

\begin{figure}[htbp!]
	\includegraphics[width=0.8\columnwidth]{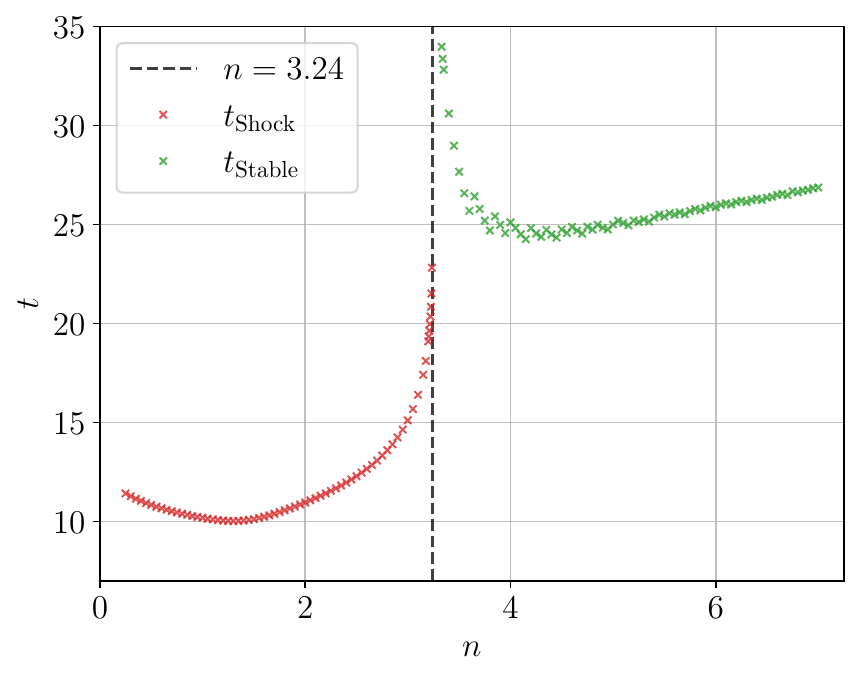}
	\caption{The transition between stable and shock forming solutions for $\eps = 0.01$, $K = 0.2$. The critical value $n^{*} \approx 3.24$ was estimated by fitting the data to \eqref{eqn:Shock_Stable_Asymptotic_Scaling}.}
\label{fig:Shock_Stable_Tranisition}
\end{figure}

\subsubsection{Fluid homogenization}
We complement the numerics with analytical estimates on a simplified model. A detailed analysis of the full system will appear in an upcoming paper \cite{FajmanMarshall:2026}.

A key mechanism for the non-linear stability is the homogenization of polytropic fluids. The analytical explanation of this effect is based on the existence of a monotone energy for the fluid variables. We consider the respective asymptotic equations ($t\gg t_0$, $|v|\ll 1$ ) near the homogeneous background spacetime where we allow for spatial inhomogeneities of $\nu=5t/4+\widetilde\nu$ and $\alpha$, while all manifestly faster decaying terms are set to zero. 
\begin{align}
\partial_t\mu
&=
-\,e^\alpha\big(\mu\,\partial_x v+v\,\partial_x\mu\big)
\;+\;\mu\Big(\frac34-\mu\Big)\label{eq-asymp-mu}\\
\partial_t v
&= v/4-e^{\alpha}v\,\partial_x v \nonumber \\
&\quad-\frac{e^{\alpha}K(1+n)}{n}\,e^{-\frac{3t}{2n}}e^{-\frac{2\widetilde\nu}{n}}\,\mu^{\frac{1}{n}-1}\,\partial_x\mu
\end{align}
In line with the numerics we assume decay of the Sobolev norms of $\tilde\nu$ and a decomposition $\alpha=t/4+\tilde\alpha$, where $\tilde\alpha(t,x)$ solves the asymptotic equation $\partial_t\tilde\alpha=3/4-\mu$. As a consequence of this and \eqref{eq-asymp-mu}, 
\begin{align*}
    e^{3t/4}\int e^{-\tilde\alpha(t,x)}(3/4-\mu)dx=\text{const}.
\end{align*}
For the $L^2$-norm of the velocity we invoke the asymptotic momentum constraint, $\partial_x\nu=-e^{-\alpha}\mu v$, which implies 
\begin{align*}
\|e^{-\tilde\alpha/2}v\|_{L^2}\leq C\|e^{-\tilde\alpha/2}\partial_xv\|_{L^2}.
\end{align*}
To measure the inhomogeneity of the fluid variables we introduce an energy, using $\nabla:=e^{\tilde\alpha}\partial_x$, by
\begin{equation}\label{eq:energy}
\begin{aligned}    
&E[\mu,v]_s:= \sum_{\ell=1}^s \int |\nabla^\ell (\mu^{1/2n})|^2 e^{-\tilde\alpha}dx\\
&\,+\frac{e^{3t/2n}}{4K(n+1)n}\int e^{2\tilde\nu/n}|\nabla^\ell v|^2e^{-\tilde\alpha}dx\\
&\,-\frac{(1+\frac{3}{4n})e^{3t/2n-t/4}}{2K(n+1)(3/4)^{\frac{1}{2n}}} \int e^{2\tilde{\nu}/n}\nabla^{\ell-1}v \cdot \nabla^\ell (\mu^{1/2n}) e^{-\tilde\alpha}dx
\end{aligned}
\end{equation}
For sufficiently large times the condition $n>3$, which is consistent with the numerical results on the critical value of the polytropic index $n^{*}$, assures that the energy is equivalent to the Sobolev norm of order $s$ of $(\mu,v)$.\\
\indent For $s\geq 2$ we obtain, by a straightforward application of the asymptotic system, the differential inequality
\begin{equation}\label{eqn:energy_identity}
\partial_t E[\mu,v]_s=-(\frac12-\frac3{4n}) E[\mu,v]_s+e^{\frac{1-3/n}4t}E[\mu,v]_s^{3/2}+\hdots,  
\end{equation}
where negligible terms are supressed. This implies decay rates of the order $e^{-(\frac12-\frac3{4n})t}$ for the energy and, in turn, for the standard Sobolev norms of the form 
\begin{equation} 
\begin{aligned}
\label{eqn:decay_estimate_fluid_norms}
    \|\mu\|_{\dot{H}^s}+e^{3t/4n}\|v\|_{\dot{H}^s}&\lesssim e^{-(\frac14-\frac3{8n})t}. 
\end{aligned}
\end{equation}

This is consistent with both numerical results on the decay of the fluid Sobolev norms (cf.~Figure \ref{fig:Fluid_H1_norms}) and on the critical polytropic index (cf.~Figure \ref{fig:Shock_Stable_Tranisition}).   
\section{Conclusions}
The results presented above provide strong numerical evidence that the Einstein–de Sitter (EdS) solution is future dynamically stable when the matter source is a polytropic fluid with polytropic index $n>3$. This behavior is in marked contrast with the corresponding Einstein–dust evolutions arising from closely comparable initial data. 
 
To the best of our knowledge, this constitutes the first numerical identification of a stable, matter-dominated attractor for the Einstein–matter system based on a physically plausible large-scale matter model relevant to the matter-dominated epoch. In particular, the stability observed here is not the outcome of a finely tuned choice of equation of state or initial data. Indeed, the polytropic model approaches dust at late times, and thus remains close to the standard dust description in the regime in which the asymptotics are assessed. Moreover, polytropic fluids are widely used in cosmological modeling, including for ranges of indices consistent with the condition $n>3$ \cite{Leubner_2005,Ostriker_2005}. We further note that stability is observed for large values of $n$, in which regime the polytropic law is arbitrarily close to the linear equations of state $p=K\rho$ commonly employed to describe isothermal fluids in cosmology. 
A detailed discussion of the physical implications will be presented in forthcoming work; nevertheless, several consequences are immediate. For sufficiently large 
$n$, the polytropic fluid homogenizes and does not form structure: in particular, no Jeans-type instability is present in the expansion-normalized variables, in the sense that the normalized energy density becomes spatially uniform while the fluid velocity decays relative to it and homogenizes. This conclusion does not, however, exclude structure formation in coupled multi-fluid settings. One may expect the polytropic component to govern the large-scale expansion with scale factor 
$a(t)=t^{2/3}$
  and to drive large-scale homogenization of the geometry, while additional components with linear equations of state may have a comparatively weaker late-time influence on the metric yet still develop non-linear structures, as suggested, for example, by analyses of shock formation in \cite{Fajman_et_al:2024}.
Finally, the most relevant consequence of the presented results is, in our opinion, the evidence for the stability of the flat FLRW model \ref{eqn:EdS}, which is a new, physically relevant non-linearly stable regime of solutions to the Einstein-Euler equations. 

\begin{acknowledgements}
\begin{center}\textbf{Acknowledgements}\end{center}
This research was funded in part by the Austrian Science Fund (FWF)
projects \textit{Matter-dominated cosmology} (10.55776/PAT7614324) and \textit{Dynamics of matter in the
decelerated epoch} (10.55776/PAT1953025). E.M. gratefully acknowledges the support of the ERC starting grant 101078061 SINGinGR, under the European Union’s Horizon Europe program for research and innovation.
\end{acknowledgements}

\bibliography{refs.bib}

\appendix

\section{Supplemental Material}
\label{app:Supplemental_Material}
In the following, we provide details on the primitive recovery process and convergence tests for our code.

\begin{figure}[htbp!]
	\includegraphics[width=0.8\columnwidth]{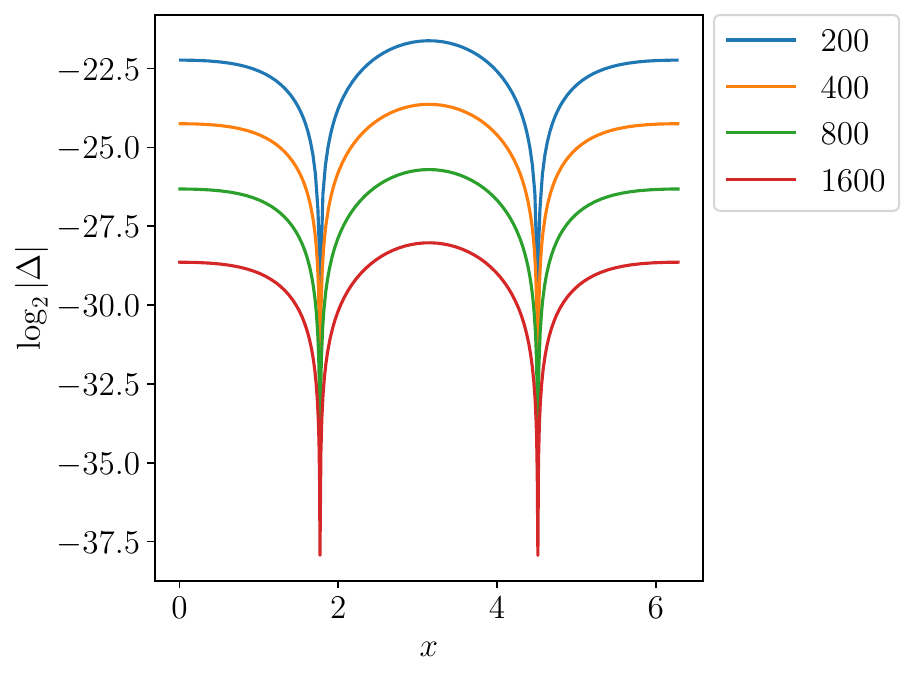}
	\caption{Convergence plot of $e^{-\alpha}\tau$ at $t\approx 5.0$. $n=5$, $K = 0.2$, $\eps = 0.01$.}
\label{fig:Convergence_tau}
\end{figure}

\subsubsection{Recovering the Primitive Variables}
Assume that the value of the expansion normalised pressure $\tilde{p}=Ke^{\frac{-2}{n}(\nu-U)}\mu^{\frac{n+1}{n}}$ is known. Then $v$ and $\mu$ can be recovered using the following formulas
\begin{equation}
\begin{aligned}
\label{eqn:PrimitiveRecovery_1}
    v = \frac{S}{\tau+\tilde{p}}, \quad
    \Gamma = \frac{1}{\sqrt{1-v^{2}}}, \quad
    \mu = \frac{\tau + \tilde{p}(1-\Gamma^{2})}{\Gamma^{2}}. 
\end{aligned}
\end{equation}
To recover the pressure itself, we solve must solve an implicit equation. We define the function $f(\tilde{p})$ by 
\begin{align}
    f(\tilde{p}) := \tilde{p} - \bar{p}
\end{align}
where
$\bar{p} := Ke^{\frac{-2}{n}(\nu-U)}\mu_{*}^{\frac{n+1}{n}}$, $\mu_{*} := \frac{\tau + \tilde{p}(1-\Gamma_{*}^{2})}{\Gamma_{*}^{2}}$, $\Gamma_{*} := \frac{1}{\sqrt{1-(\frac{S}{\tau+\tilde{p}})^{2}}}$. The derivative $f^{\prime}(\tilde{p})$ is given by
\begin{align}
f^{\prime}(\tilde{p}) &=  1 - \frac{Ke^{\frac{-2}{n}(\nu-U)}\frac{n+1}{n}\mu_{*}^{\frac{1}{n}}S^{2}}{(\tau+\tilde{p})^{2}}.
\end{align}
In our code, we solve for $\tilde{p}$ using the Newton-Raphson method.

\begin{table}[htbp!]
\centering
\begin{tabular}{|l|l|l|l|l|}
\hline
\multicolumn{1}{|l}{$N$} &  \multicolumn{1}{|l|}{$\|u_{N} - u_{3200}\|_{L^{2}}$} & \multicolumn{1}{|l|}{Order} & \multicolumn{1}{|l|}{$\|u_{N} - u_{3200}\|_{L^{\infty}}$} & \multicolumn{1}{|l|}{Order} \\ \hline
200& $4.64 \times 10^{-7}$ & - &  $3.11\times10^{-7}$ & - \\ 
400 & $1.15 \times 10^{-7}$ & $2.02$  & $7.69 \times 10^{-8}$ & $2.02$ \\  
800 & $2.73 \times 10^{-8}$ & $2.07$ & $1.83\times10^{-8}$& $2.07$ \\ 
1600 & $5.46 \times 10^{-9}$  & $2.32$ & $3.66\times 10^{-9}$& $2.32$  \\ 
\hline                                
\end{tabular}
\caption{Numerical error and convergence order for $e^{-\alpha}\tau$ at $t \approx 5.0$. $n=5$, $K = 0.2$, $\eps = 0.01$.}
\label{table:tau_convergence_error}
\end{table}

\subsubsection{Convergence Tests}
We have tested the convergence of our code using resolutions of $N \in \{200, 400, 800, 1600, 3200\}$. We estimate the numerical error $\Delta$ by taking the difference of each solution and the (appropriately restricted) highest resolution ($N=3200$) run. A convergence plot for $e^{-\alpha}\tau$ is shown in Figure \ref{fig:Convergence_tau}, see also Table \ref{table:tau_convergence_error}, from which the second order convergence is clear. Similarly, the $L^{2}$ norm of the momentum constraint violation is shown in Figure \ref{fig:Convergence_MomentumConstraint} which also demonstrates the expected second order convergence.

\begin{figure}[htbp!]
	\includegraphics[width=0.8\columnwidth]{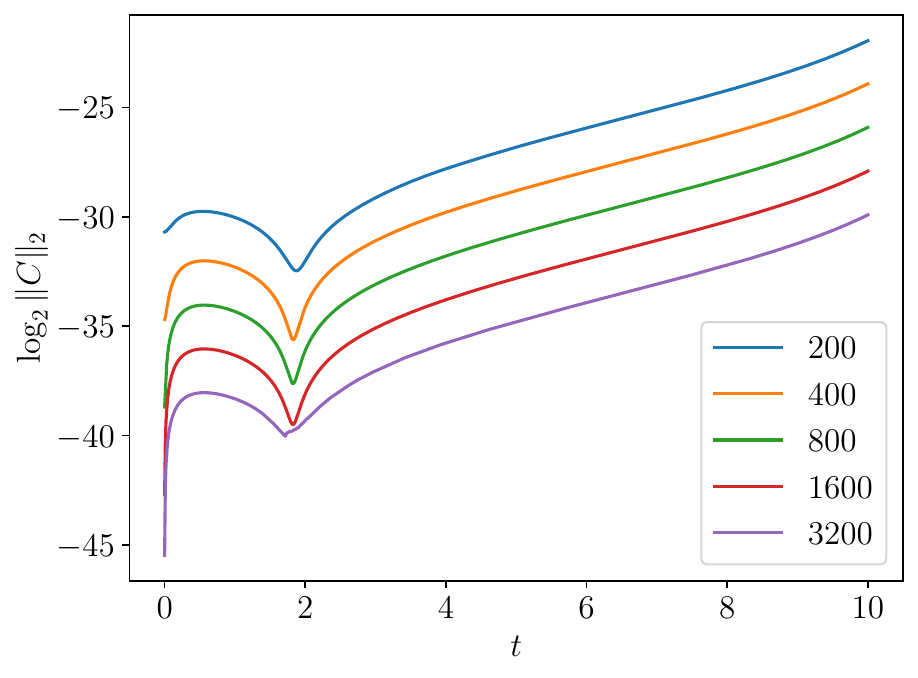}
	\caption{$L^{2}$ norm of the momentum constraint violation. $n=5$, $K = 0.2$, $\eps = 0.01$.}
\label{fig:Convergence_MomentumConstraint}
\end{figure}

\end{document}